\documentclass{aa}
\usepackage{psfig}

\def\be{\begin{equation}}
\def\ee{\end{equation}}
\def\bea{\begin{eqnarray}}
\def\eea{\end{eqnarray}}

\begin{document}
\thesaurus{02       
          (12.12.1; 
           12.04.3; 
           11.04.1; 
           11.19.7; 
           03.13.6)}

\title{Spatial Structure and Periodicity in the Universe}

\author{Jos\'e A. Gonz\'alez \and Hernando Quevedo \and
 Marcelo Salgado \and Daniel Sudarsky} 
\offprints{M.~Salgado}

\institute{Instituto de Ciencias Nucleares \\
Universidad Nacional Aut\'onoma de M\'exico \\
A. P. 70--543, M\'exico, D. F. 04510,  M\'exico\\
{\em e-mail : cervera@nuclecu.unam.mx, quevedo@nuclecu.unam.mx \\ 
marcelo@nuclecu.unam.mx, sudarsky@nuclecu.unam.mx}}

\date{Accepted August 2000}

\maketitle

\begin{abstract} 
We analyze the possibility that the spatial periodicity 
of 128$h^{-1}$ Mpc  in the galaxy count 
number observed recently in deep pencil-beam surveys 
could be explained  by the {\it mere}
existence of structure  with the appropriate scale in the 
galaxy distribution in the Universe. 
We simulate a universe where the
distribution of galaxies has an intrinsic length scale and
then 
investigate the probability of observing in a given direction
a spatial periodicity similar to that observed in the pencil-beam 
surveys of the galactic polar regions. A
statistical analysis shows that this probability is of the order
of $10^{-8}
$, a value which excludes the afore mentioned explanation.
This result contrasts with the estimates based on Voronoi foam models
which yield probabilities of the order $10^{-1} - 10^{-2}$.
The conclusion that emerges from these studies put together with
the present one, is that although they can be obtained in some models,
the presence of an intrinsic scale in the structure is not sufficient 
to achieve reasonable probabilities for observation of the detected
periodicities.  
\end{abstract}

\section{Introduction}

The cosmic microwave 
background radiation shows a remarkable isotropic structure
of the order
$\delta T/T \approx 10^{-5}$, which is taken as
an evidence for the smoothness of the mass density of the 
Universe on the largest scales comparable to the Hubble
radius (Wilkinson 1987), $H_0^{-1} \approx 3000h^{-1}$ Mpc 
($H_0 = 100 h $ km s$^{-1}$ Mpc$^{-1}$). On the other
hand, the existence of galaxies and clusters of galaxies 
as well as the (larger than unity) amplitude of the 
galaxy-galaxy correlation function (Peebles 1980) indicate
that the Universe possesses a highly irregular 
structure on scales less than 10$h^{-1}$ Mpc.
 
On intermediate scales, ranging from about 30$h^{-1}$ Mpc
to around 2000$h^{-1}$ Mpc, the existing data do not allow a 
conclusive interpretation about the mass density distribution
of the Universe. Nevertheless, there are a number of 
different observations suggesting that the structure on 
these scales is quite varied. The most interesting of these
observations include the voids with sizes greater than 30$h^{-1}$ Mpc 
observed in
the distribution of bright galaxies 
(Kirschner et al. 1981; de Lapparent et al. 1986) the (larger
than unity) amplitude of the cluster-cluster correlation function
on scales up to 30$h^{-1}$ Mpc, the peculiar velocity field in
our local vicinity on a scale of about 50$h^{-1}$ Mpc, and the 
``Great Wall" at a distance of approximately 100$h^{-1}$ Mpc 
(Geller \& Huchra 1980). Also,  noteworthy is the analysis of the 
 spatial distribution of superclusters on the intermediate 
scales (Einasto et al. 1997). For this analysis a sample of more than
200 superclusters (systems with more than 8 clusters where the 
distances between nearest-neighbors among member clusters do not
exceed 24$h^{-1}$ Mpc) was used, where the estimated redshift 
of simple members reaches up to $z=0.12$ (around 1/4 of the 
redshift reached in the pencil-beam galactic surveys). These 
three-dimensional data of the distribution of superclusters 
has been used to calculate the correlation function as well
as the power spectrum. In the analysis of the 
power spectrum it was found that it is characterized by a peak 
with a wavelength of ($120\pm 15)h^{-1}$ Mpc.

But probably the most puzzling discovery concerns the recent 
observations which show an  
apparent periodicity in the distribution of redshifts in 
pencil-beam surveys of the north and south galactic polar regions 
(Broadhurst et al. 1990). These surveys extend to a redshift of around
$z\approx 0.5$ in both directions, and there is a strong evidence
for a periodicity of 128$h^{-1}$ Mpc which extends for over 13
periods. Although this result was considered initially as a 
statistical anomaly  
(Kaiser \& Peacock 1991), later and more detailed 
studies (Kopylov et al. 
1988; Mo et al. 1992; Fetisova et al. 1993; Einasto \& Gramann 1993; 
Szalay et al. 1993; Einasto et al. 1997) 
provided new evidence for the alluded periodicity on the same scales.
 
If this periodicity would be established in most directions, a naive 
interpretation would imply that galaxies and clusters are 
distributed on concentric shells centered around our galaxy, 
clearly a blow to our cosmological conceptions. 
Another more plausible explanation is based upon the idea 
that the observed periodicity is just a visual effect induced
by a periodic oscillation of the gravitational constant 
(Morikawa 1990; Hill, Steinhardt \& Turner 1990; Salgado, 
Quevedo \& Sudarsky 1996, 1997; Quevedo, Salgado \& Sudarsky 1997; 
Sudarsky, Quevedo \& Salgado 1998), and that
the mass density of the Universe on intermediate scales remains
homogeneous and isotropic as on the largest scales.
This explanation requires the introduction of a 
time-dependent scalar field into the model 
(the oscillating $G$ model) which is non-minimally coupled to 
gravity. 

In view of the remarkable conclusions that would derive 
from any of these explanations it seems imperative to consider whether 
the observations are likely to occur in the absence of these scenarios, 
and can be the result of chance augmented with the assumption of 
the presence  of structure at the appropriate scale but with
 an otherwise random distribution.
We will study the plausibility of this latter explanation inspired
on the idea,  suggested 
by Szalay et al. (1993), that  
the structure of the Universe on these scales consists of a 
regular network of superclusters and voids with a step size of
the order of the wavelength discovered in the power spectrum. 
The  specific structure that is studied by Van de Weygaert (1991)
and Subba Rao and Szalay (1992)
resembles a collection  of honeycombs or a 
three-dimensional chess-board. Such structures known as 
Voronoi foams result in a reasonable probability for the observation
of a periodicity similar to that in the actual observations.

The aim of this work is to investigate whether it is sufficient 
to have a universe with an intrinsic length scale in its large scale 
structure in order to explain the observation of the periodicity
mentioned above.
In particular, the present work can be seen as a test of the 
suggestion that a hypothetical ``built-in scale in the initial 
spectrum'' arising from double inflation models and supported by
cosmic microwave background radiation data (Einasto, 1997), could be
the explanation of the above mentioned periodicity. 
Therefore, we will consider a model that does not introduce any other
large scale pattern besides the existence of the above mentioned 
characteristic scale.

Needless is to say that neither our model nor the Voronoi foam models are 
anything more than ``toy'' models as they lack, for example, a proven
dynamical scenario for their formation starting from reasonable initial 
data. 
We want to construct a simple realization of such a model.
 We will do this by assuming that
 the galaxies are evenly distributed over 
the surface of three-dimensional spheres with an average radius
comparable with the appropriate scale, with the spheres themselves
randomly distributed in the Universe.  
In a universe filled by these spheres we choose a 
point as the ``observational site'' and perform a statistical analysis to 
investigate the probability of observing a periodic distribution with the
 characteristics of those reported in (Broadhurst et al. 1990)
 in a particular direction. We obtain an upper bound for the value of
such probability of the order 
$10^{-8}$, a value that obviously excludes the alluded 
structure as an acceptable explanation of the observations.
The assumption
 that the galaxies are homogeneously 
distributed on the surface of the spheres is not expected to affect the
main result of our numerical analysis in the  direction
that would invalidate this upper bound.
 
\section{The model}

We have taken  the radii of the spheres to 
have a Gaussian distribution centered around the mean value   
of 80$h^{-1}$ Mpc and analyzed the results for different choices 
of the standard deviation $\sigma$ of the distribution 
(namely with $\sigma=5,15,30h^{-1}$ Mpc).
We have implemented a code that simulates the universe as a 
three-dimensional 
volume of $(5000\,h^{-1} {\rm Mpc})^3$ filled by these spheres subjected  
to the condition that the spheres do not intersect with
each other. In figure 1, a two-dimensional section of this volume is 
depicted. 

\begin{figure}
\centerline{\psfig{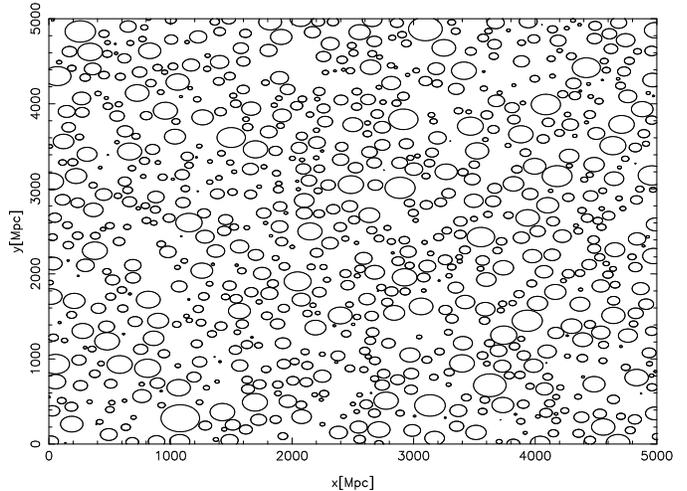}}
\caption[]{
Two-dimensional section of a universe filled with a random smearing of
three-spheres with a Gaussian distribution in radii. A constraint is imposed 
in order that the spheres do not intersect with each other. Notice that 
the distortion of the circles into ellipses is only due to the 
difference of scales on the x$-$y axes.\label{f:fig1}}
\end{figure}

We emphasize that the above condition of no-sphere overlapping  
and the Gaussian distribution on the spheres radii imply that the 
numerical process of filling the toy universe stalls after 
a certain number of spheres. 
After that stage, the inclusion of more spheres demands a very long
computing time. The adequate number of spheres needed to fill the universe 
is not fixed in advance, but the important point 
is that such a number must be so that the 
correlation function of the matter in the universe
exhibits a characteristic scale, which 
is the most important aspect in the scenario that we are testing. 
We have confirmed that once correlations are present
the results become insensitive to a further increase
in the number of  spheres. 

In order to check the appearance of scale, we
have computed a two-point correlation function defined in the
following way. Let $N$ be the total number of spheres 
in the toy universe, and let $n_i(r)$ be the number 
of sphere centers contained between two concentric 
shells with radii $r$ and
$r + \delta r$ around the center $i$ 
(with $\delta r \sim 10$ Mpc $h^{-1}$). The number density of 
sphere centers within  $r$ and $r + \delta r$ is then given by
\begin{equation}
C(r) = {1\over N}\sum_{i=1}^N {n_i(r)\over r^2}.
\end{equation} 
This is to be constant if correlations between 
neighbors at all scales are absent. It is customary to extract from this, the 
excess density number by defining the two-point correlation function as the 
difference between the above density distribution and the mean 
density of sphere-center number:
\begin{equation}
\xi(r) = C(r)- <C>.
\end{equation}
More specifically, $<C>$ corresponds to the mean of 
$C(r)$ at scales for which $C(r)$ is nearly constant (i.e., 
$\xi\sim 0$ at those scales). 
If the Universe has a characteristic length scale we expect it will be
 evident in the plot of the correlation function.

The results are shown in figure 2 with a typical number 
$N\approx 4\times 10^4$. We can appreciate 
the presence of a characteristic scale at distances which are comparable 
to the mean value of the spheres radii, and stress 
that such scale is already present when the spheres fill a 
volume of only about 25\% of the total volume of the toy universe.  
We do not expect any drastic changes in the qualitative as well as 
in the quantitative behavior
of the correlation function if we increase the number
 of spheres in the universe. Moreover, as shown in figure 2,  
the resulting length scale is almost independent of the 
value of $\sigma$ considered.

\begin{figure}
\centerline{\psfig{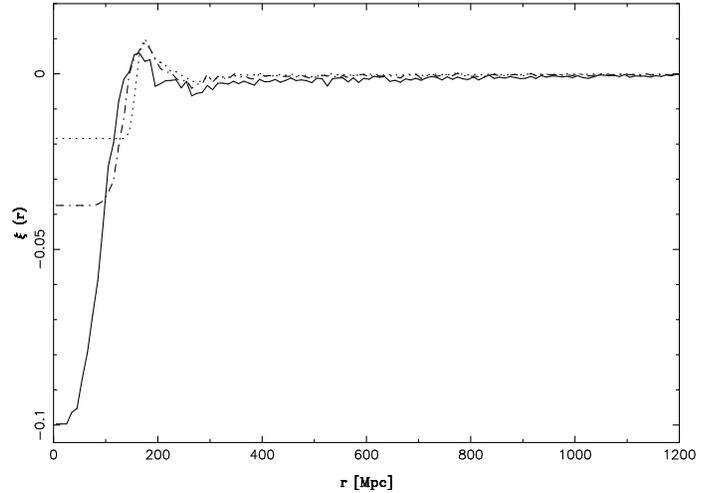}}
\caption[]{Two-point correlation function of the matter content as a function of 
the distance. The dotted, solid and dash-dotted lines corresponds to 
the values $\sigma=5$, $\sigma=30$ and $\sigma=15$ respectively. 
Notice the presence of correlation at scales of $\sim$160-180 Mpc and 
anticorrelations at shorter scales.\label{f:fig2} }
\end{figure}

\bigskip

Now we will estimate the probability of  observing  in a specific direction
a periodicity like the one seen in the real observations.
To this end we note that the  power spectrum of the 
observations in Szalay et al. (1993) revealed a peak at a frequency 
corresponding to 
128$h^{-1}$ Mpc  which contained 23\% of the total integrated power.
Thus we need to 
estimate the probability of  observing  in a specific direction
a matter distribution with a  power spectrum containing a
peak with approximately the same percentage 
 of the total power.

The ``observational site'' has been chosen to be a vertex of the 
box which in fact represents an octant of a larger toy-universe. 
This particular choice allows to consider the longest possible extent for the 
line of sight in each direction, but it does not affect
the result of the statistical analysis which is performed for
about 10$^5$ directions and for ``pencil-beams'' extending 
approximately to 5000$h^{-1}$ Mpc.   

 In this model each direction corresponds to 
a numerical pencil-beam, that is, a straight 
line which intersects several spheres, each 
intersection thus representing high-density spots. The ensemble of the  
distance separations between the spots results thus in a 
one-dimensional 
distribution of matter along the line of sight, which we 
represented mathematically by an ensemble of step functions of 
width $10$ Mpc $h^{-1}$ with height normalized to unity centered 
on each intersection of the line of sight with the surface of a 
sphere. We have also consider the effect of replacing step functions
by Gaussian functions and showed that this does not change the probability
estimates 
beyond the statistical error bars. 
Finally, 
we have performed a Fourier analysis of this distribution and 
confronted the resulting power spectrum with the observational one. 
Figure 3 shows the resulting power spectrum  of the matter 
distribution in a representative direction. 
We note that there is a substantial power at high frequencies, a feature 
associated with the use of step functions that is slightly suppressed
by the use of Gaussian functions.

In order to compare with the real observations we look in each 
direction for 
the peak containing most of the power.
We then calculate the percentage $\Pi$ of the power-content within the peak:
\be
\Pi = {{\hbox{Power content within the peak}}
\over {\hbox{Total power}} }\times 100  \ .
\ee
As we mentioned, this procedure has been applied to 10$^5$ different 
directions starting from the specific site described above,  
obtaining representative peaks with values $\Pi$ at different frequencies
in each case. The distribution of the number of peaks with a given $\Pi$ 
is plotted in figure 4 as a function of its corresponding wavelength. 
We note that most of the highest peaks cluster around a specific wavelength 
of $\sim 40$ Mpc $h^{-1}$, 
however the power contained in most of them represents barely 
2\% of the corresponding total integrated power (see for instance 
the figure 3). 

\begin{figure}
\centerline{\psfig{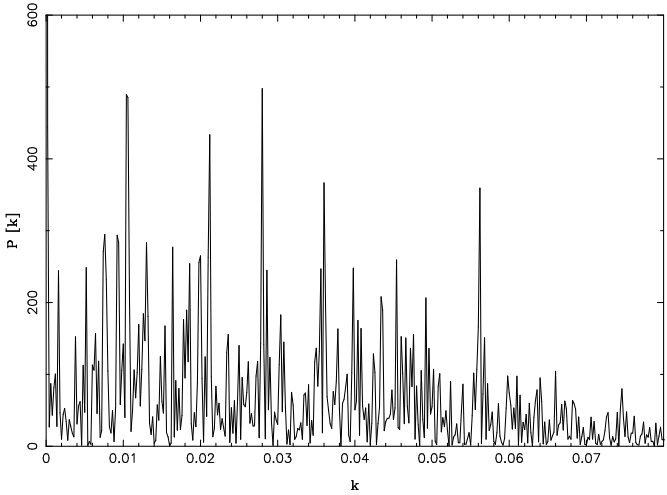}}
\caption[]{Example of a power spectrum (Fourier analysis) of the resulting 
distribution of matter along a line of sight.\label{f:fig3}}
\end{figure}

\begin{figure}
\centerline{\psfig{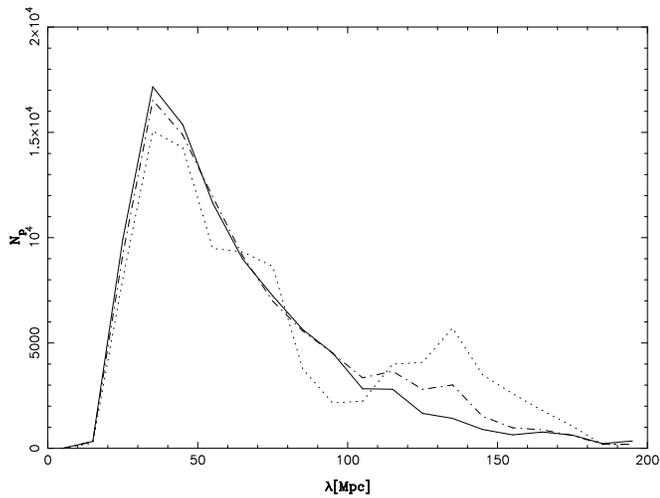}}
\caption[]{Distribution of the number of 
peaks with a given $\Pi$ (maximum relative power content)
 associated to each direction of 
sight as a function of their corresponding 
wavelength in the Fourier space. 
The dotted, solid and dash-dotted lines corresponds to 
the values $\sigma=5$, $\sigma=30$ and $\sigma=15$ respectively.
\label{f:fig4}}
\end{figure}

Since one can define the probability of observing a periodicity in a
certain direction as the number of directions in which the periodicity 
is present over the total number of directions, one should 
analyze an adequate sample in which one observes a sufficient number of
 cases with the desired periodicity. 
Since the sample of 10$^5$ directions showed no 
periodicity with a peak having the 
required power $\Pi\sim 23\%$,  
one should be tempted to give only an upper bound 
of 10$^{-5}$ for the probability of observing that value. 
This is not the best upper bound that can be obtained 
and it is only determined by the finiteness of the
 computing time. We have been able nevertheless to improve on this bound
by the following procedure: We have fitted our data  for
the probability $p$ of a peak with a given percentage $\Pi$ of the power 
 as a function of that percentage
to  various analytical expressions 
and found that this function is very well fitted
by a power law:
\be\label{fit}
p(\Pi) = c \Pi^{\alpha} \,\,\,\,\,\\
\ee
where $c$ and $\alpha$ are constants with values depending 
on the chosen $\sigma$ (see table 1).  
This allows us to extrapolate the probability 
of finding the observed periodicity with $\Pi=23\%$ in a given direction.  
 
\begin{table}
\caption[]{ \label{t:ranges}
Parameters of the power-law (\ref{fit}) for the probability of 
finding a spatial periodicity in a given direction. Here the 
statistical uncertainties of data are taken into account for determining 
the ranges of the parameters.}
\begin{flushleft}
\begin{tabular}{lccc}
\hline\noalign{\smallskip}
$\sigma$ & $c$ & $\alpha$ & $\chi^2$    \\
\noalign{\smallskip}
\hline\noalign{\smallskip}
30           & 2.446564 $\times$ $10^6$ $\pm$ 1.722128 $\times$ $10^6$
 & $-7.30$ $\pm$ 0.20 & 0.401 \\
\hline
15           & 2.390874 $\times$ $10^6$ $\pm$ 1.682928 $\times$ $10^6$  
& $-7.22$ $\pm$ 0.20 & 0.377 \\
\hline
5           & 1.3196748 $\times$ $10^7$ $\pm$ 1.0245539 $\times$ $10^7$  & 
$-8.81$ $\pm$ 0.31 & 2.468 \\
\noalign{\smallskip}
\hline
\end{tabular}
\end{flushleft}
\end{table}

An example of this law for $\sigma=30$ Mpc is shown in figure 5.  
A similar fit was made for the cases with other values
of $\sigma$ with the remarkable result that the power law is 
practically unaffected. This is a strong 
evidence for the robustness of our results. 

\begin{figure}
\centerline{\psfig{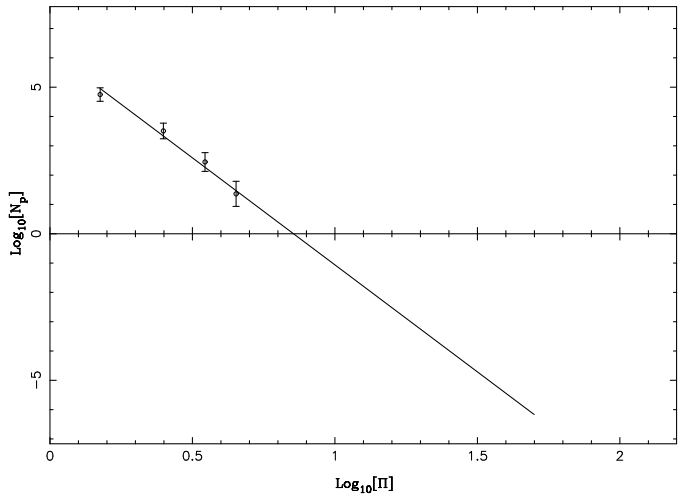}}
\caption[]{Power law probability (not normalized) of obtaining a 
peak with a corresponding value $\Pi$. 
Note that the higher the relative power content of the representative 
peak the lower the number of directions of sight where such a 
peaks is present. In particular the probability of observing a peak 
containing 23\% of the total integrated power is $\sim 10^{-8}$. 
\label{f:fig5}}
\end{figure}

We also stress that the data-set fitted to the above power law 
has values of $\chi^2$, which according to the degrees of freedom of 
the data ($\sim 2$), indicates the good quality of the fits (see table 1). 
This allows to make confident predictions   
for the probability of observing the value of 
$\Pi= 23\%$ which corresponds to  
the observations of Szalay et al. (1993).
 From (\ref{fit}) one easily concludes  
that the value of $p(23\%)$ ranges in the interval  
$10^{-8}-10^{-10}$. This interval 
is determined by taking into account the statistical uncertainties of the fits.
This implies that the probability of obtaining a peak with the 
observational power-content is at most 10$^{-8}$. Needless to say that this 
result completely discards the possibility of detecting  the observed
periodicity in a universe with this kind of structure. 
\bigskip

As we have stressed,  
we have performed a similar analysis replacing the step functions
by Gaussian functions
 with the result that the upper bound for the 
probability $p$ remains essentially unaffected and all the values of
$p$ remain within the interval determined by the statistical 
uncertainties. 
\bigskip

It should be emphasized that in this work we are considering the possibility
of detecting periodicity in a particular
 direction under the {\it sole} assumption
that there exists an intrinsic length scale in the galaxy distribution. 
Our results show that this probability is completely negligible. However,
this result could drastically be changed within a different model which 
would be characterized not only by a length scale but also by the presence
of a specific correlation in the spatial distribution of galaxies. A 
concrete example of a model with this type of additional structure 
has  
been investigated  (Van de Weygaert, 1991; 
Subba Rao \& Szalay, 1992). In these 
works, dynamical Voronoi foams (Icke \& Van de Weygaert, 1987;
Van de Weygaert \& Icke, 1989) with a controlled amount of 
randomness are proposed as describing the spatial 
galactic distribution. The probability for finding periodicity 
in these models depends on the amount of randomness and can
drastically be increased from 3\% up to 15\%. This probability 
can reach even higher values (10\% to 40\%) 
in the case of a regular
lattice (Kurki-Suonio et al., 1990). In the extreme and naive 
model of galaxies distributed on the surfaces of concentric 
shells, one obviously would obtain a 100\% probability for an 
observer at the center. We want to stress that the models that
are more easily reconciled with acceptable cosmological scenarios
are those that introduce a priori the least amount of scale order,
and such that the order introduced corresponds to 
the smallest possible scale. 
In this sense the models that introduce just a characteristic length
scale and no other large scale order should be considered as constituting
the most viable explanations for the observations.

The conclusion and the main result of our analysis is that the 
existence of a specific scale in the distribution of mass density of the
universe, as showed by a two-point correlation function, does not by itself
provide a 
satisfactory explanation for the spatial periodicity of 
128$h^{-1}$ Mpc  
in the galaxy distribution of deep pencil-beam surveys as it 
results in a negligible probability for such observation
to be obtained in a particular direction. This result seems to 
force us again to look for alternative explanations for the 
observations of periodicity in deep pencil-beam surveys. 
On the other hand, it might be interesting to consider variations 
of this type of model (for exemple, changing the intersection rule) to  
identify the features that lead to an enhaced probability for obtaining the 
desired periodicity.

\begin{acknowledgements}
This work has been supported by DGAPA-UNAM, grant Nr. 121298 
and CONACYT, Mexico, grant Nr. 3567-E. The authors thank the 
Department of Supercomputing of DGSCA-UNAM for facilitating to us 
the use of the Cray-Origin-2000.
\end{acknowledgements}

\end{document}